\documentclass[amsmath,notitlepage,5p,times,twocolumn]{elsarticle}
\usepackage{amssymb}
\usepackage{amsmath}
\usepackage{slashed}
\usepackage{graphicx}
\usepackage{xcolor}
\usepackage{titlesec}
\usepackage{subfig}
\usepackage{multirow}
\usepackage{hhline}
\usepackage{lipsum}
\usepackage{hyperref}

\newcommand\TeV{\mbox{\,TeV}}
\newcommand\GeV{\mbox{\,GeV}}
\newcommand\pb{\mbox{\,pb}}
\newcommand\nb{\mbox{\,nb}}
\newcommand\mub{\,\mu\mbox{b}}
\newcommand\mb{\mbox{\,mb}}
\newcommand\Dmeson{\mbox{D}}
\newcommand\sigEff{\sigma_{\mbox{eff}}}
\newcommand\feeddown{\mbox{f.-d.}}

\begin{document}
\title{Production of associated $\chi_b$ and open charm at the LHC}

\author[aa,bb]{A. K. Likhoded}
\ead{Anatolii.Likhoded@ihep.ru}

\author[aa]{A. V. Luchinsky}
\ead{Alexey.Luchinsky@ihep.ru}

\author[aa]{S. V. Poslavsky}
\ead{stvlpos@mail.ru}
\address[aa]{Institute for High Energy Physics NRC ``Kurchatov Institute'', 142281 Protvino, Moscow Region, Russia}
\address[bb]{Moscow Institute of Physics and Technology, 141700 Dolgoprudny, Moscow Region, Russia}

\begin{abstract}
In the present paper we study the production of $\chi_b + c\bar c$ at the LHC within single parton scattering approach. A special attention payed to the feed-down from $\chi_b$ states to the associated $\Upsilon + c\bar c$ production, which was recently studied by the LHCb. We have found that this feed-down is about percents of the total cross section seen in the experiment. It is shown that the shapes of the differential distributions are almost same for single and double parton scattering approaches except the azimuthal asymmetry, which is the most distinguishing feature of the latter one. We conclude that the precise study of the single parton scattering contributions is necessary for the correct isolation of the double parton scattering role.
\end{abstract}
%\keywords{heavy quarkonia production}

\maketitle

\section{Introduction}

% http://xxx.lanl.gov/pdf/1006.3846v2.pdf  http://arxiv.org/pdf/1007.3095.pdf  !!! COLOR OCTET !!!
Multiple production of heavy quarks attracts a great interest during recent years, and with the launch of the LHC a huge data sample on these processes became available. Some of such processes, like production of $B_c$ mesons, can certainly be described within standard single parton scattering (SPS) approach. On contrary, theoretical predictions obtained for the processes like double $J/\psi$, associated $J/\psi$ + open charm and double open charm production often underestimate experimental cross sections. This is often explained by the fact that other channels, such as double parton scattering (DPS) can give a contribution. While in the case of double $J/\psi$ production it is still possible to reconcile the SPS with the observed cross sections \cite{Berezhnoy:2011xy, Baranov:2011zz,Sun:2014gca}, in the open charm sector the observed cross sections \cite{Aaij:2012dz} are larger significantly than the SPS predictions and better fit into the DPS picture \cite{Berezhnoy:1998aa, PhysRevD.73.074021,Lansberg:2008gk,refId0}.

It is worth to mention, that the DPS approach is very attractive by its simplicity; the cross section within DPS can be simply obtained via:
$$
\sigma^{AB} = \frac{\sigma^A \times \sigma^B}{\sigEff},
$$
where $\sigma^{AB}$ is a cross section of paired production of particles $A$ and $B$, $\sigma^{A,B}$ are the cross sections of single production, and $\sigEff$ is some ``effective'' cross section which is determined experimentally. Despite the simplicity, we cannot say that we fully understand this mechanism, e.g.~what is the physical sense of the dimensional non-perturbative parameter $\sigEff$ and how it is related to the fundamental parameters of the QCD. Moreover, the experimental value obtained for the $\sigEff$ differs significantly from one experiment to another varying in the range $(2.2 \div 20)\mb$ \cite{Abazov:2015fbl, Akesson:1986iv, Aaij:2015wpa, Abazov:2014qba, Abe:1993rv, Chatrchyan:2013xxa}. Having DPS model as a ``bad defined'' in some sense, the precise theoretical and experimental input is crucial.

The processes of double production of heavy quarkonia with different flavour i.e.~$\Upsilon$ and $J/\psi$ can be extremely helpful in the understanding of the underlying mechanism of heavy quarkonia production. This is because the direct production $p p \to \Upsilon + J/\psi + X$ is forbidden in the leading order  $\sim\alpha_s^4$ within the SPS approach, so one can expect that other channels should come to the fore. In the recent paper \cite{Likhoded:2015zna} we have calculated the process of $P$-wave quarkonia production $p p \to \chi_b + \chi_c + X$ and the corresponding feed-down to the $\Upsilon + J/\psi$, which we have found is about 2\% of the DPS prediction for this process. On the other hand, our rough estimations presented ibid give that the NLO contribution increase the SPS prediction in several  times, which leaves much less space for the DPS.

Recently the LHCb performed measurement of the associated production of $\Upsilon$ and open charm \cite{Aaij:2015wpa}. The direct production of $\Upsilon + c\bar c$ was studied in \cite{Berezhnoy:2015jga} both within SPS and DPS approach. It was found that the SPS mechanism gives contribution of the order of one percent. Thus, we may expect a significant feed-down contribution from the $\chi_b + c\bar c$ and from the NLO $\Upsilon + c\bar c$ processes. The results of \cite{Aaij:2015wpa} suggest the DPS approach is valid in both description of total cross sections and cross section distributions.

\begin{figure}[t]
\centerline{\includegraphics[width=0.8\columnwidth]{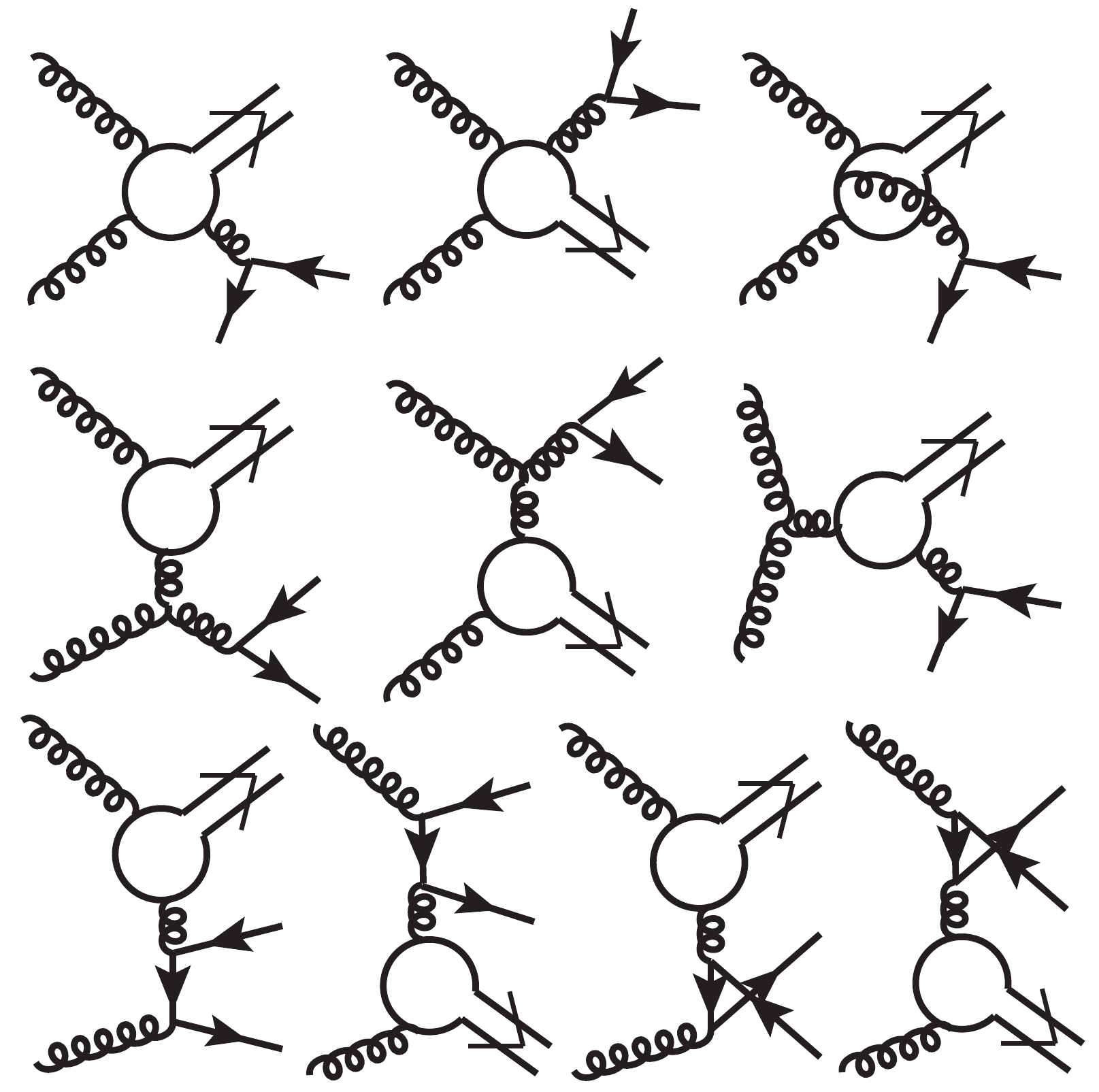}}
\caption{\label{fig_diagrams} Feynman diagrams contributing to the process \eqref{eq_hard_process} at the leading order.}
\end{figure}

The rest of our paper is organized as follows. In the next section we consider matrix elements and cross sections of the partonic reaction $gg\to\chi_{bJ}+c\bar{c}$. In Section III various distributions and total cross sections of the hadronic reaction $pp\to\chi_{bJ}+D\bar{D}+X$ are presented. Short discussion of the obtained results is given in the last section. Technical details of the calculations can be found in the Appendix.

\section{Parton level}

The relevant process on the parton level at the LHC energies is gluon fusion:
\begin{equation}
\label{eq_hard_process}
g + g \to \chi_{bJ} + c + \bar c.
\end{equation}
Corresponding Feynman diagrams are shown in Fig.~\ref{fig_diagrams}. According to the NRQCD factorization theorem \cite{Bodwin:1994jh}, the cross section of $\chi_{bJ}$ production can be expressed in a series in powers of relative quark-antiquark velocity $v$:
\begin{equation}
\label{eq_nrqcd_series}
d \hat \sigma( \chi_J ) = \sum_{n} \mathcal O^{\chi_{bJ}}\, ([b \bar b]_{n}) d \hat \sigma([b \bar b]_{n}),
\end{equation}
where $n$ denotes a set of spin $S$, angular momentum $L$ and color quantum numbers, and parameters $\mathcal O^{\chi_{bJ}}\, ([b \bar b]_{n})$  are determined by the non-perturbative matrix elements responsible for $[b \bar b]_{n}$ pair hadronization into observable state with possible emission of soft gluons (so called $\text{E}1$ chromo-electric, $\text{M}1$ chromo-magnetic, $\text{E}1\times\text{E}1$ transitions and so on). Since the relative velocity of $b\bar b$ pair in bottomonium is small ($v^2 \approx 0.1$), series \eqref{eq_nrqcd_series} has a good convergence.

The leading terms in \eqref{eq_nrqcd_series} come from color-singlet $[b \bar b] (^3P_J^{[1]})$ and $\text{E}1$ color-octet $[b \bar b] (^3S_1^{[8]})$ contributions which are of order $O(v^2)$. Next to leading corrections $O(v^4)$ come from $\text{M}1$ chromo-magnetic $[b \bar b] (^1P_1^{[8]})$ and $\text{E}1\times\text{E}1$ double chromo-electric $[b \bar b] (^3P_J^{[8]})$ transitions which are of order $O(v^4)$. As it is was shown in our previous works on a single $\chi_b$ production \cite{Likhoded:2014kfa,Likhoded:2013aya,Likhoded:2012hw}, all color-octet contributions are negligibly small and become important only when considering ratio of cross sections with different total spin (e.g. $\sigma(\chi_{b2})/\sigma(\chi_{b1})$) or at very high $p_T$ region which is not yet accessible in the experiment. Thus, we will take into account only dominant color-singlet contribution in \eqref{eq_nrqcd_series}. In the latter case, the corresponding non-perturbative matrix element can be expressed in terms of phenomenological non-relativistic wave function of the meson:
$$
\mathcal O^{\chi_{bJ}}\, ([b \bar b] (^3P_J^{[1]})) = \frac{3}{4\pi} \left( 2 J + 1\right) \left|R'(0)\right|^2,
$$
where $R(r)$ is a radial part of meson's wave function.

\begin{figure*}[t]
\begin{tabular*}{\textwidth}{c @{\extracolsep{\fill}} c}
\includegraphics[width=0.45\textwidth]{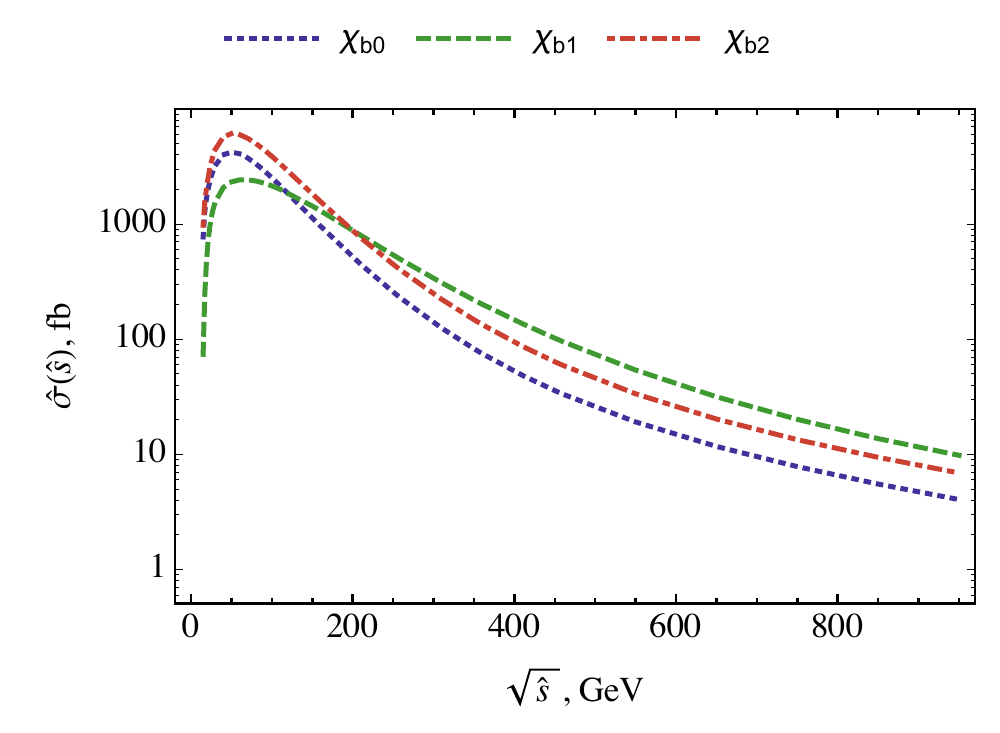} & \includegraphics[width=0.45\textwidth]{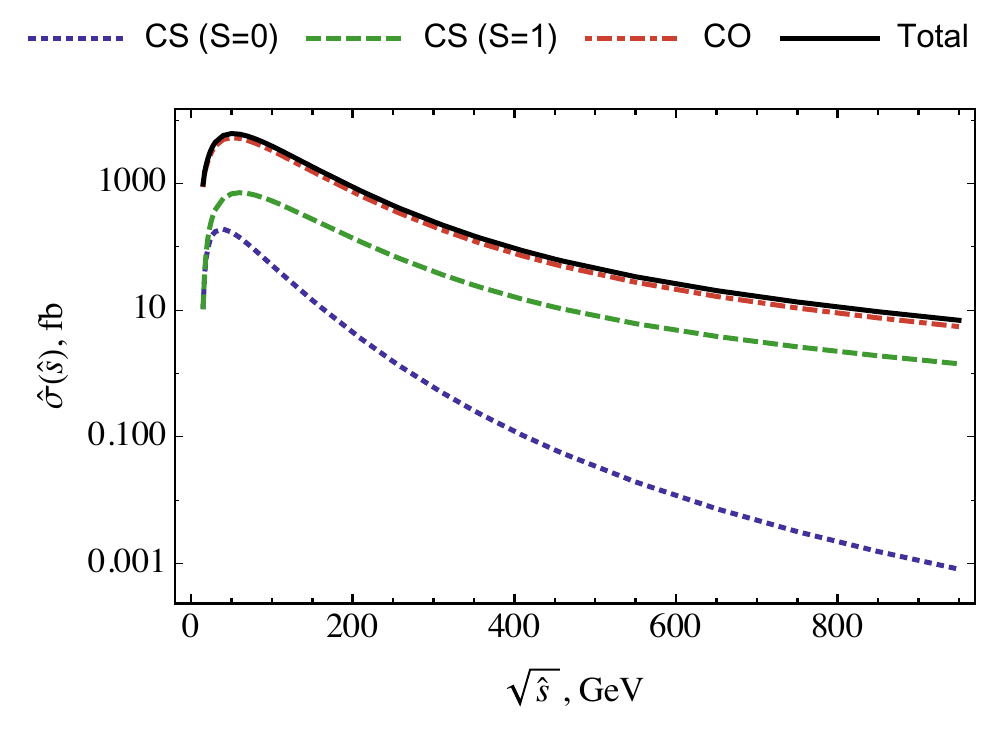}
\end{tabular*}
\caption{\label{fig_hard_cs_contribs} 
(left) Dependence of the hard cross sections on the partons energy:  dotted, dashed and dot-dashed curves correspond to the processes $gg\to\chi_{b0,1,2} + c\bar c$ respectively. 
(right) Contributions from different $c\bar c$-channels to the total cross section of $gg\to\chi_{b2} + c\bar c$: dotted, dashed and dot-dashed curves correspond to the color singlet $S=0$, $S=1$ and color octet states of $c\bar c$ pair; solid curve is a sum. The cross sections in the figure are calculated with $\alpha_S = 0.25$. }
\end{figure*}

\begin{figure*}[t]
\begin{tabular*}{\textwidth}{c @{\extracolsep{\fill}} c}
\includegraphics[width=0.45\textwidth]{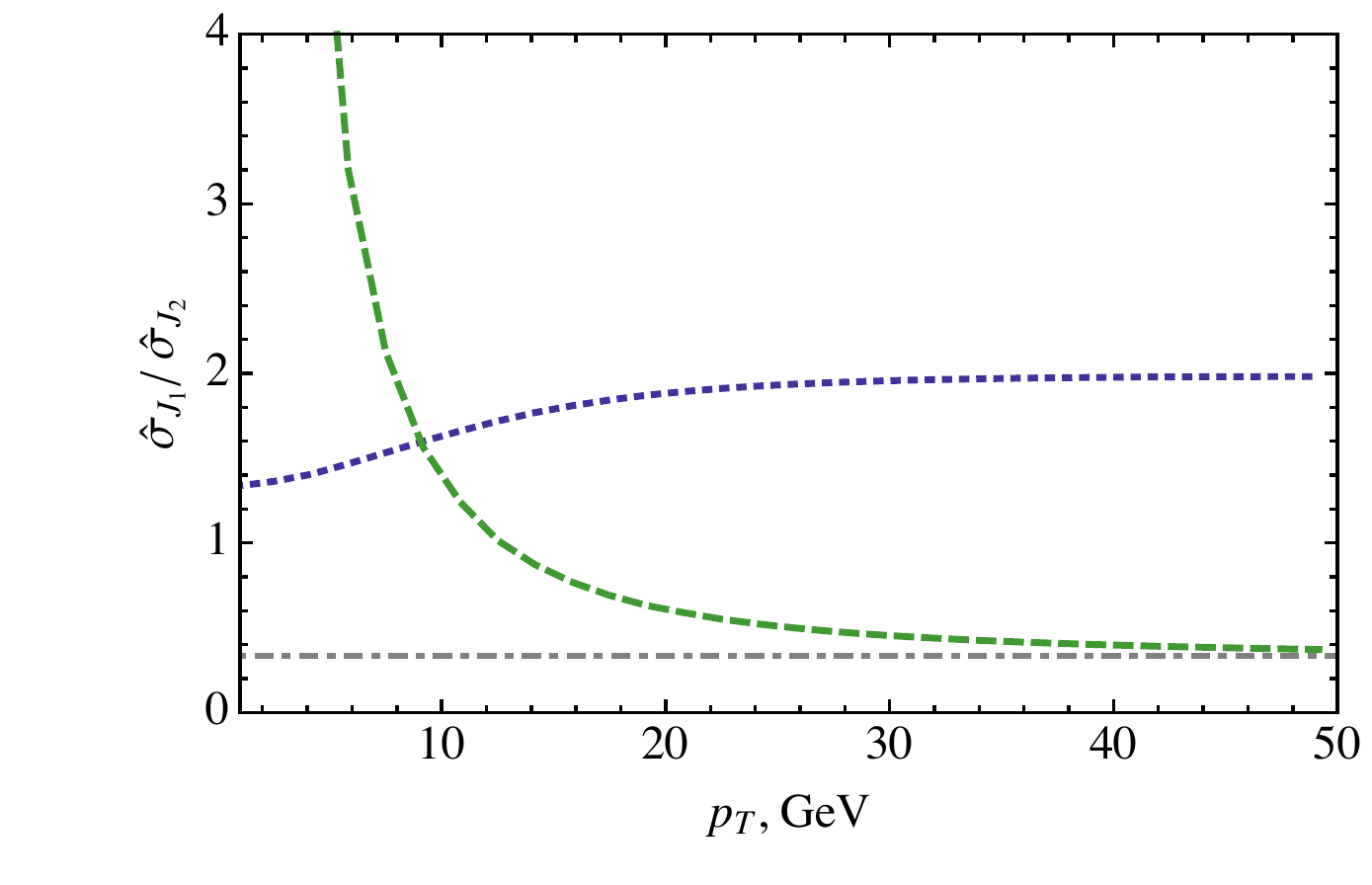} & \includegraphics[width=0.45\textwidth]{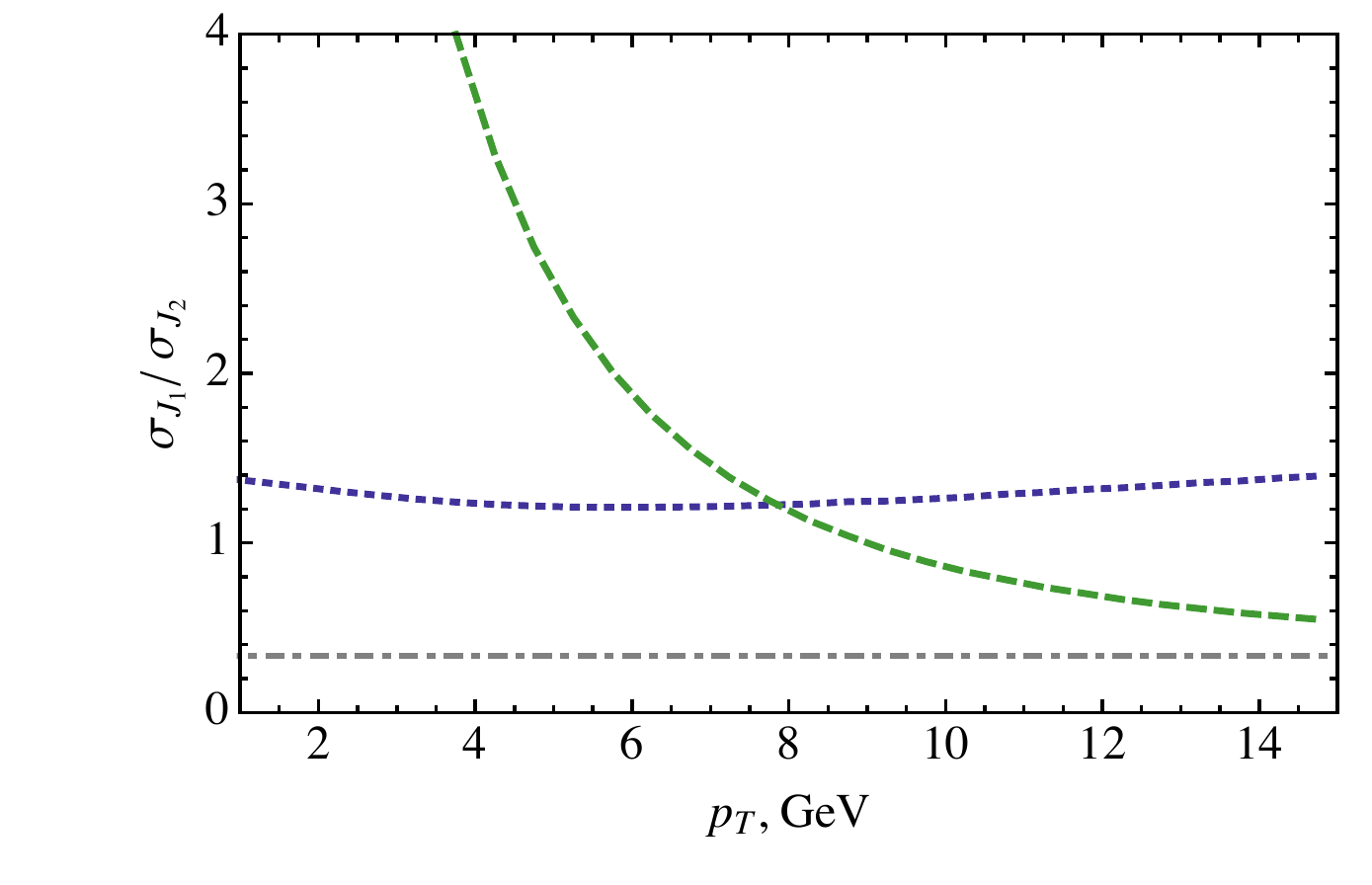}
\end{tabular*}
\caption{\label{fig_hard_cs_pt} 
Transverse momentum distribution for the cross section ratio $\sigma(\chi_{b2})/\sigma(\chi_{b0})$ (dotted) and $\sigma(\chi_{b2})/\sigma(\chi_{b1})$ (dashed). The left figure for the partonic cross sections evaluated at $\sqrt{\hat s} = 150 \GeV$, right -- for the hadronic cross sections at $\sqrt{s} = 8 \TeV$. The constant dot-dashed line corresponds to the $1/3$.
}
\end{figure*}

The details of the hard cross sections calculation can be found in \ref{app_calc}. Fig.~\ref{fig_hard_cs_contribs} (left) shows the dependence of the hard cross sections for different $J$-states on the total energy of initial partons. We have also calculated contributions for different quantum numbers of $c\bar c$-pair: when $c\bar c$ in color singlet (in this case only the last four diagrams from Fig.~\ref{fig_diagrams} give a non zero contribution) with spin $S_{c\bar c} = 0$ or $1$, and when it is in the color octet combination. Fig.~\ref{fig_hard_cs_contribs} (right) shows these contributions for $\chi_{b2}$ in the final state (for other $\chi_b$ states pictures are similar). From this figure, it is clear that the dominant part of the cross section comes from colored $c\bar c$ combinations.

There is one interesting property of the cross sections. The dependence on the $p_T$ at $p_T \ll M_{\chi_b}$ and $p_T \gg M_{\chi_b}$ is the same as in the case of single $\chi_b$ production $gg\to \chi_b + g$. On the one hand, it is clear that at the small values of $p_T$ we have for the ratio:
$$
\left. \frac{d \hat \sigma(\chi_{b2})/d p_T}{d \hat \sigma(\chi_{b0})/d p_T}\right|_{p_T \ll M_{\chi_b}} \to
 \frac{5}{1}\times  \frac{\Gamma(\chi_{b2} \to gg )}{\Gamma(\chi_{b0} \to gg) } = \frac{4}{3},
$$
where the first factor comes from $(2 J + 1)$ factor. On the other hand at the large values of $p_T$ the explicit calculation shows that 
$$
\left. \frac{d \hat \sigma(\chi_{b2})/d p_T}{d \hat \sigma(\chi_{b1})/d p_T}\right|_{p_T \gg M_{\chi_b}} \to \frac{1}{3},
$$
which is the same behaviour as for the single $\chi_b$ production (under the assumption that color octet contributions are negligible, see \cite{Likhoded:2014kfa} for the details). Fig.~\ref{fig_hard_cs_pt} (left) shows the dependence of the ratios of hard cross sections on $p_T$. It is clear, that the same behaviour will hold for the hadronic reactions as well (Fig.~\ref{fig_hard_cs_pt} (right)).

\section{Hadron level}

The cross section of hadron process within single parton scattering approach can be written as:
\begin{equation}
\label{eq_hadrons_cs}
d \sigma = \int dx_1 dx_2 f_g(x_1; Q^2) f_g(x_2; Q^2) d \hat \sigma,
\end{equation}
where $f_g(x; Q^2)$ are gluon distributions at the scale $Q^2$. We used CT10 PDF sets \cite{Dulat:2015mca} with the LHAPDF interface \cite{Buckley:2014ana}. Both strong coupling and PDFs were taken at $\mu^2 = Q^2 = M_{\chi_b}^2 + 2m_c^2 + p^2_{T\chi_b}$ scale. We have performed calculation of the cross sections with different kinematical cuts; here we present results for the LHCb kinematical region which is $2 < y(\chi_b, \Dmeson^{0,+}) < 4.5$ at $\sqrt{s} = 8$ and $13$ TeV; results for other kinematical regions (ATLAS, CMS, D0) are available on request. 

Since the $\chi_b$ mesons are detected via its radiative decays to $\Upsilon(1S)$ state, we simulated these radiative transitions in our estimations. On the other hand, we have neglected nontrivial fragmentation function of the $c$-quark $c \to \Dmeson^{0,+}$, assuming that all $c$-quarks hadronize in D mesons with hundred percent probability and that the momentum of the final D-meson is almost same as for $c$-quark; account for nontrivial fragmentation will slightly change the numerical values but not the main results.

For the $\sqrt{s} = 8$ TeV we have found the following cross sections:
\begin{eqnarray*}
 \sigma(\chi_{b0} + c \bar c)_{\sqrt{s}=8\TeV}\, &=& \,115.7 \, \pm \, \phantom{0}7.8 \pb, \\
 \sigma(\chi_{b1} + c \bar c)_{\sqrt{s}=8\TeV}\,  &=& \,\phantom{0}43.5 \, \pm \, \phantom{0}0.1 \pb, \\
 \sigma(\chi_{b2} + c \bar c)_{\sqrt{s}=8\TeV}\,  &=& \,152.6 \, \pm \, 18.7 \pb.
\end{eqnarray*}
For the $\sqrt{s} = 13$ TeV:
\begin{eqnarray*}
 \sigma(\chi_{b0} + c \bar c)_{\sqrt{s}=13\TeV}\, &=& \, 251.0 \, \pm \, 5.8 \pb, \\
 \sigma(\chi_{b1} + c \bar c)_{\sqrt{s}=13\TeV}\,  &=& \,\phantom{0}99.8 \, \pm \, 0.1 \pb, \\
 \sigma(\chi_{b2} + c \bar c)_{\sqrt{s}=13\TeV}\,  &=& \,321.2 \, \pm \, 5.9 \pb,
\end{eqnarray*}
which is nearly two times larger than at $\sqrt{s} = 8$ TeV. The errors in above formulas are due to the Monte-Carlo which we used for integration in \eqref{eq_hadrons_cs}. Cross sections are presented with the assumption that $ |R'(0)|^2 \approx 1 \GeV^5 $, which is in a general agreement with the potential models for $\chi_b(1P)$ and $\chi_b(2P)$ states (see e.g. Tab.~1 in \cite{Likhoded:2012hw}). 

The feed-down to the $\Upsilon(1S)$ from the $\chi_{bJ}(nP)$ states is determined by the following formula:
\begin{multline} 
\label{eq_ups_from_chi}
 \sigma_{\feeddown}^{\chi_b(nP)}(\Upsilon + c \bar c) =\\
= \sum_{J = 0}^2 \mathcal  B [{\chi_{bJ}(nP) \to \Upsilon(1S) + X}]  \times  \sigma(\chi_{bJ}(nP)  + c \bar c),
\end{multline}
where a possible double transition $\chi_b(2P) \to \Upsilon(2S) \gamma \to \Upsilon(1S) + X$ included. Summing contributions from $n=1,2$ we obtain:
\begin{eqnarray*}
\sigma_{\feeddown }(\Upsilon+c\bar c)_{\sqrt{s}=\phantom{0}8\TeV} \, &=& \, \phantom{0}72.2 \, \pm \, 5.9 \pb,\\
\sigma_{\feeddown }(\Upsilon+c\bar c)_{\sqrt{s}=13\TeV } \, &=& \,156.3 \, \pm \, 3.0 \pb.
\end{eqnarray*}
The total $\Upsilon + c \bar c$ cross section recently measured by the LHCb \cite{Aaij:2015wpa} at $\sqrt{s}=8\TeV$ is the following:
\begin{equation*}
\sigma^{\mbox{\scriptsize LHCb}}_{\sqrt{s}=8\TeV }(\Upsilon + \Dmeson^{0,+}) \, = \, 13.2 \pm 1.8 \,\mbox{(stat)} \pm 0.6 \,\mbox{(syst)} \nb.
\end{equation*}
Thus, the feed-down from $\chi_{bJ}$ states produced via SPS to the $\Upsilon(1S) + c \bar c$ production is about 0.6\%. 

On the other hand, recall that the process under consideration is of order $\alpha_S^4$. We have used rather big scale $\mu^2 = M_{\chi_b}^2 + 2 m_c^2 + p^2_{T\chi_b}$ at which $\alpha_S$ is small; choice $\mu = M_{\chi_b}^2$ will raise the cross section in approximately 5 times resulting in 3\% feed-down contribution to the associative $\Upsilon(1S) + c \bar c$ production. We have calculated the total cross sections at different scales, namely at $Q^2/2$ and at $2\times Q^2$, where $Q^2$ is a value defined at the beginning of the section and found that the results vary in about two times. Another possible source of the undercount is an uncertainty in the wave function of the meson: for example, in the case of prompt $\chi_c$ production we showed in \cite{Likhoded:2014kfa} that in order to fit existing data we have to raise the $|R'(0)|$ in several times from its phenomenological value. In total, we conclude that at the current level of accuracy we found that the feed-down contribution from $\chi_b$ states to the associative $\Upsilon + c\bar c$ production is at least about percents:
$$
\left. \left. {\sigma_{\feeddown }^{\mbox{\scriptsize SPS}}} \right/ {\sigma^{\mbox{\scriptsize LHCb}}}\right|_{\sqrt{s}=8\TeV } = \, (0.6 \div 1.5)\%
$$

Let's now estimate the DPS predictions for the total cross sections $\chi_b + c\bar c$. For the $\sigma(c\bar c)$ we take LHCb results from \cite{Aaij:2013mga}:
$$
\sigma^{\mbox{\scriptsize LHCb}}(c\bar c)_{\sqrt{s}=7\TeV} = 1419 \,\pm\, 12 \, \pm \,116 \,  \pm \, 65 \, \mub,
$$
and roughly scale it with a factor $8/7$ to obtain estimation at $\sqrt{s} = 8\TeV$. The total cross section for the $\Upsilon(1S)$ originating from $\chi_b$ mesons we take from our previous paper \cite{Likhoded:2012hw}:
$$
\sigma_{\feeddown }(\Upsilon(1S))_{\sqrt{s}=8\TeV} = 30.5 \nb.
$$
It is interesting to note that this prediction is in a good agreement with the experimental results: LHCb measured both total $\Upsilon(1S)$ production rate \cite{LHCb:2012aa,Aaij:2015awa} and a fraction of $\Upsilon(1S)$ originating from $\chi_b$ decays \cite{Aaij:2012se,Aaij:2014caa}. Combining this, the obtained experimental value for the $\chi_b$ feed-down is:
$$
\sigma^{\mbox{\scriptsize LHCb}}_{\feeddown }(\Upsilon(1S))_{\sqrt{s}=8\TeV} = 27 \, \pm \, 8 \nb.
$$
Taking $\sigEff \approx 18 \mb$ obtained in \cite{Aaij:2015wpa} we have for the DPS:
\begin{gather*}
\sigma_{\feeddown }^{\mbox{\scriptsize DPS} }(\Upsilon(1S)+c\bar c)_{\sqrt{s}=8\TeV} = \frac{\sigma_{\feeddown }(\Upsilon(1S)) \times  \sigma^{\mbox{\scriptsize LHCb}}(c\bar c)}{\sigEff}\\
 \phantom{\sigma_{\feeddown }^{\mbox{\scriptsize DPS} }(\Upsilon(1S)+c\bar c)_{\sqrt{s}=8\TeV}} 
\approx 2.75 \, \pm \, 0.39 \nb.
\end{gather*}
This shows that the SPS cross section is roughly about 2.6\% of the DPS one at $\sqrt{s} = 8\TeV$. Applying similar considerations about uncertainties coming from $\alpha_S$ and $R'(0)$ we may conclude that this value may be increased significantly. 

Another interesting thing that one can see is that the SPS prediction at $\sqrt{s} = 13\TeV$ is roughly two times larger than at $\sqrt{s} = 8\TeV$. While the shape of the hard cross section is the same, it is intuitively clear that the reason of the cross section growth is in the gluon densities which grows at small $x$-values that become accessible at higher energies. This is also confirmed by the LHCb for the $J/\psi$ production \cite{Aaij:2015rla}, where they found $\mathcal R^{\mbox{\scriptsize}}_{13/8} \approx 2$. Thus, me may assume that such behaviour is rather universal, i.e. valid for other SPS dominated processes as well:
$$
\mathcal R^{\mbox{\scriptsize SPS}}_{13/8} = \left. {\sigma^{\mbox{\scriptsize SPS}}_{\sqrt{s}=13\TeV}}\right/ {\sigma^{\mbox{\scriptsize SPS}}_{\sqrt{s}=8\TeV}}\approx 2,
$$
in what follows that for the DPS dominated process we will have:
$$
\mathcal R^{\mbox{\scriptsize DPS}}_{13/8} = \left.{\sigma^{\mbox{\scriptsize DPS}}_{\sqrt{s}=13\TeV}} \right/ {\sigma^{\mbox{\scriptsize DPS}}_{\sqrt{s}=8\TeV}}\approx 4
$$
(under the assumption that $\sigEff$ is a constant). Thus, the ratio of the cross sections at $\sqrt{s} = 8 \TeV$ and $\sqrt{s} = 13 \TeV$ will be another clear test of the DPS mechanism.

Let's now move focus on the differential cross sections and correlations between final particles. While we have found that the total SPS cross section for $\chi_b + c\bar c$ is about several percents of the DPS one, these correlations can give an idea of the differences in these two  mechanisms at a more detailed level. For these considerations we will use the normalized distributions
$
(1/ \sigma)\, d \sigma / d v
$
and apply the same conventions as in \cite{Aaij:2015wpa}. The experimental data from LHCb for the total $\Upsilon(1S) + c\bar c$ production will be used as a reference.

\begin{figure*}[t]
\begin{tabular*}{\textwidth}{c @{\extracolsep{\fill}} c}
\includegraphics[width=0.45\textwidth]{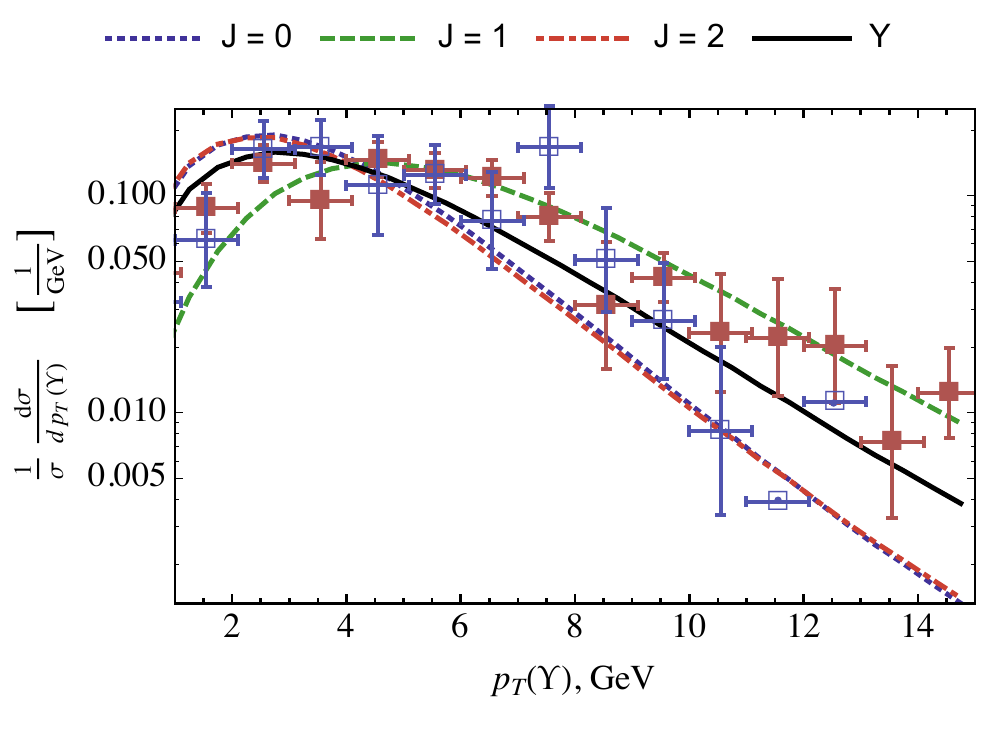} & \includegraphics[width=0.45\textwidth]{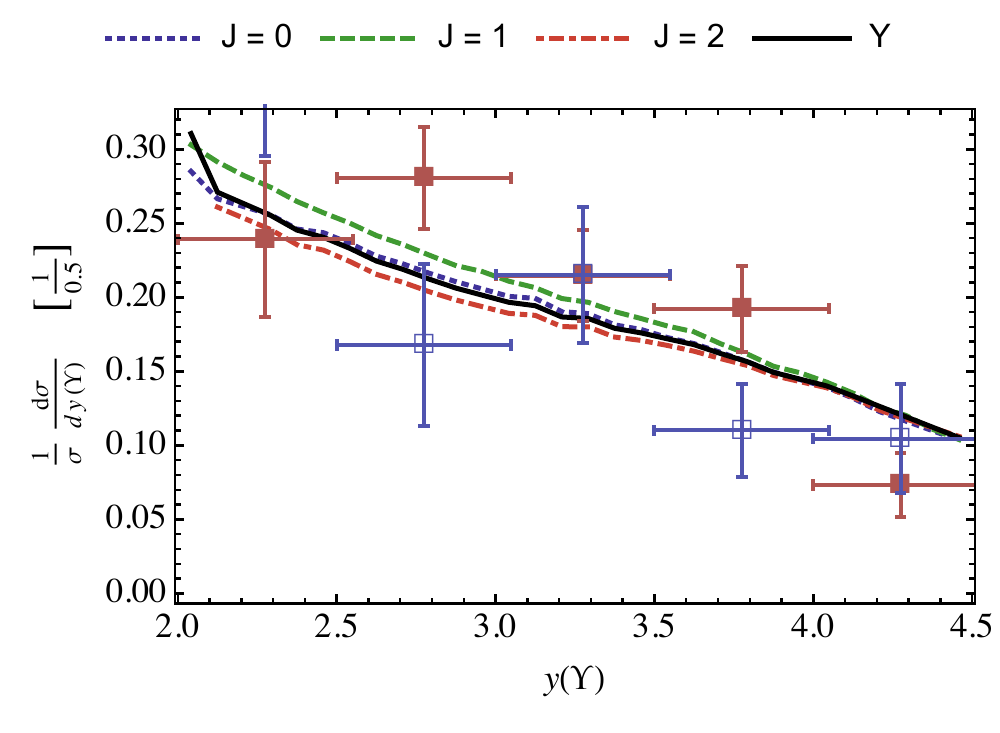} 
\end{tabular*}
\caption{ \label{fig_distr_for_Y_1} Normalized cross section distributions for transverse momentum $p_T$ and rapidity $y$ of $\chi_b$ meson obtained within the SPS model. Dotted, dashed and dot-dashed curves correspond to the $\chi_{b0}$, $\chi_{b1}$ and $\chi_{b2}$ states respectively, solid curve --- to the $\Upsilon(1S)$ state produced via radiative $\chi_b$ decays \eqref{eq_ups_from_chi}. Data points for the total $\Upsilon(1S) + \Dmeson^{0,+}$ production are taken from LHCb \cite{Aaij:2015wpa}: filled rectangles for the $\Upsilon(1S) \Dmeson^0$ final state and open rectangles -- for the $\Upsilon(1S) \Dmeson^+$.}
\end{figure*}

\begin{figure*}[t]
\begin{tabular*}{\textwidth}{c @{\extracolsep{\fill}} c}
\includegraphics[width=0.45\textwidth]{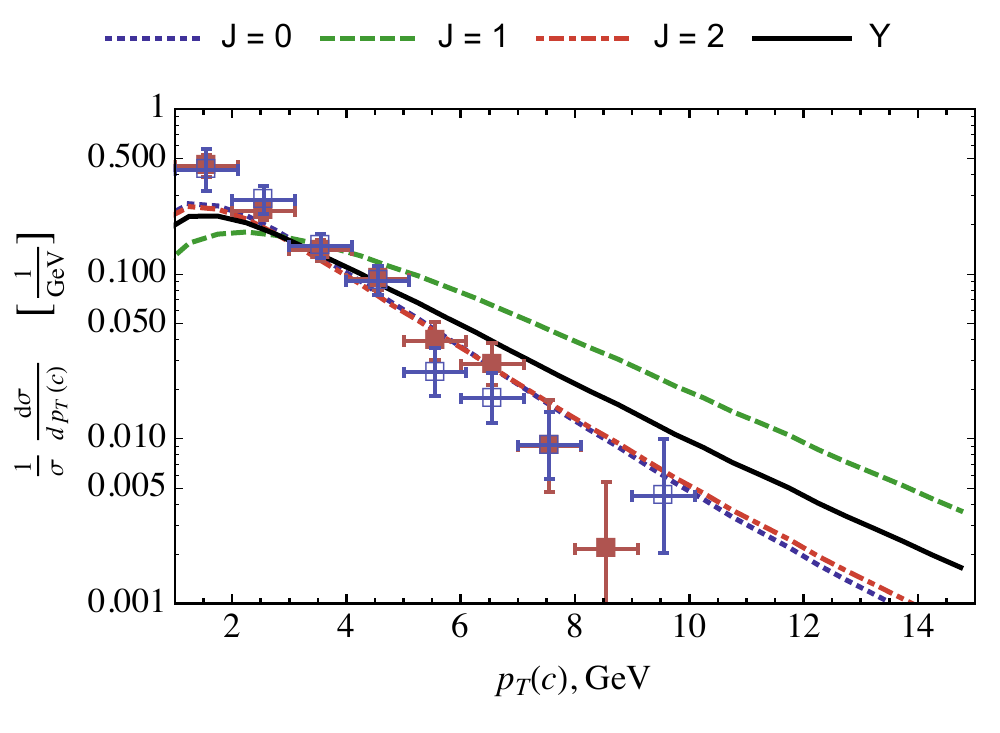}& \includegraphics[width=0.45\textwidth]{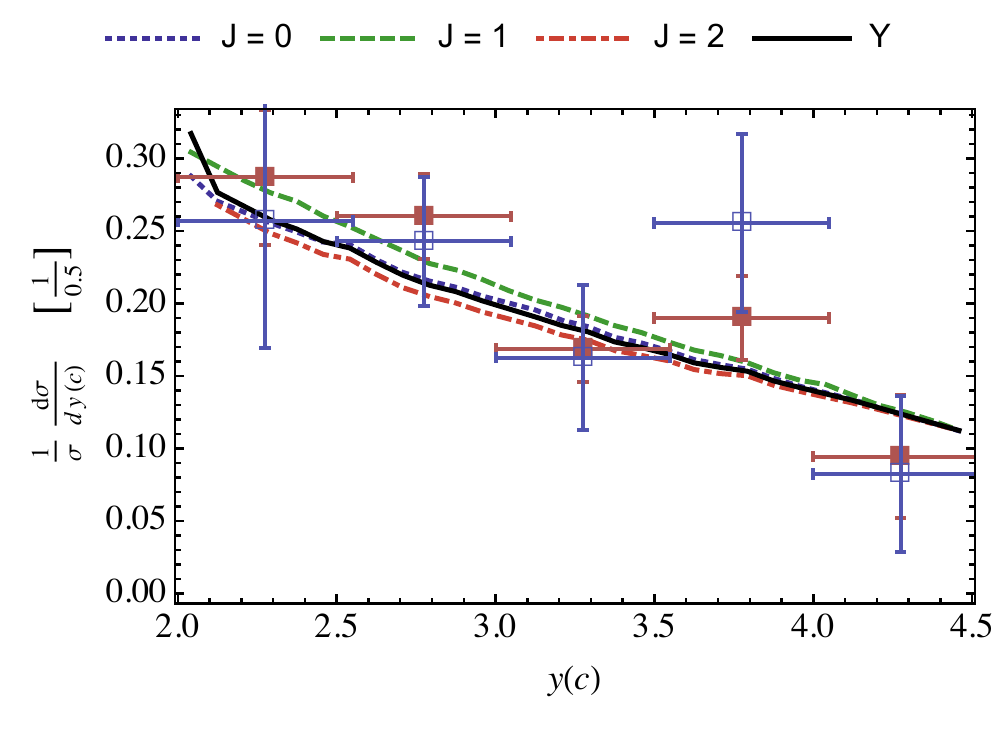}\\
\end{tabular*}
\caption{ \label{fig_distr_for_Y_2} Normalized cross section distributions for transverse momentum and rapidity of $c$-quark obtained within the SPS model. The definitions are the same as for Fig.~\ref{fig_distr_for_Y_1}.}
\end{figure*}

\begin{figure*}[t]
\begin{tabular*}{\textwidth}{c @{\extracolsep{\fill}} c}
\includegraphics[width=0.45\textwidth]{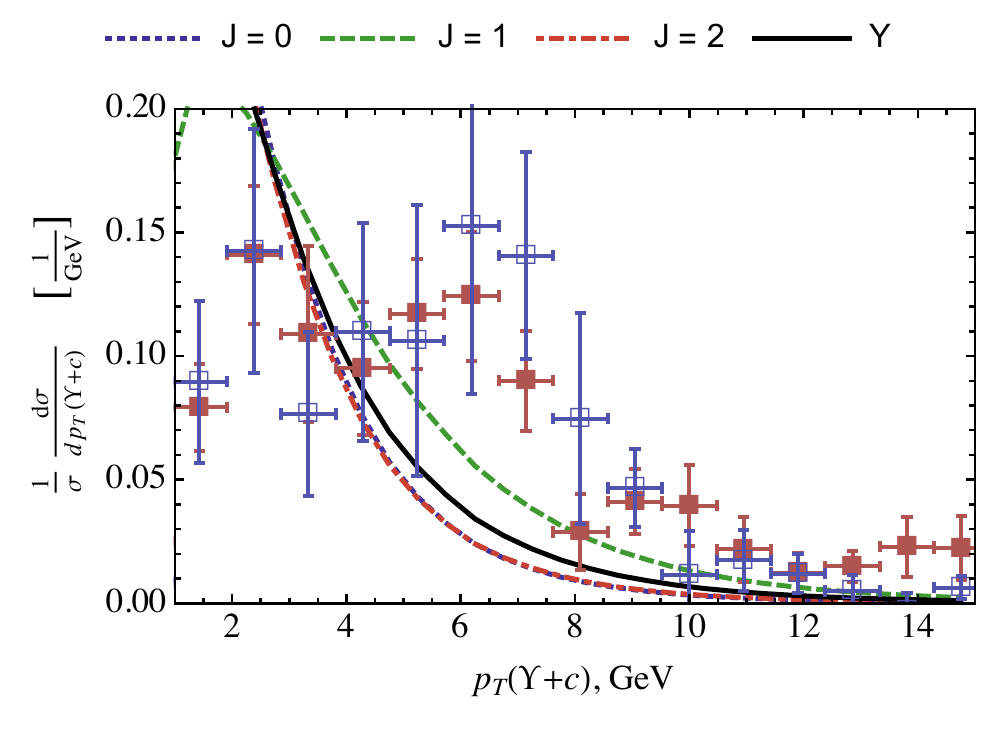} & \includegraphics[width=0.45\textwidth]{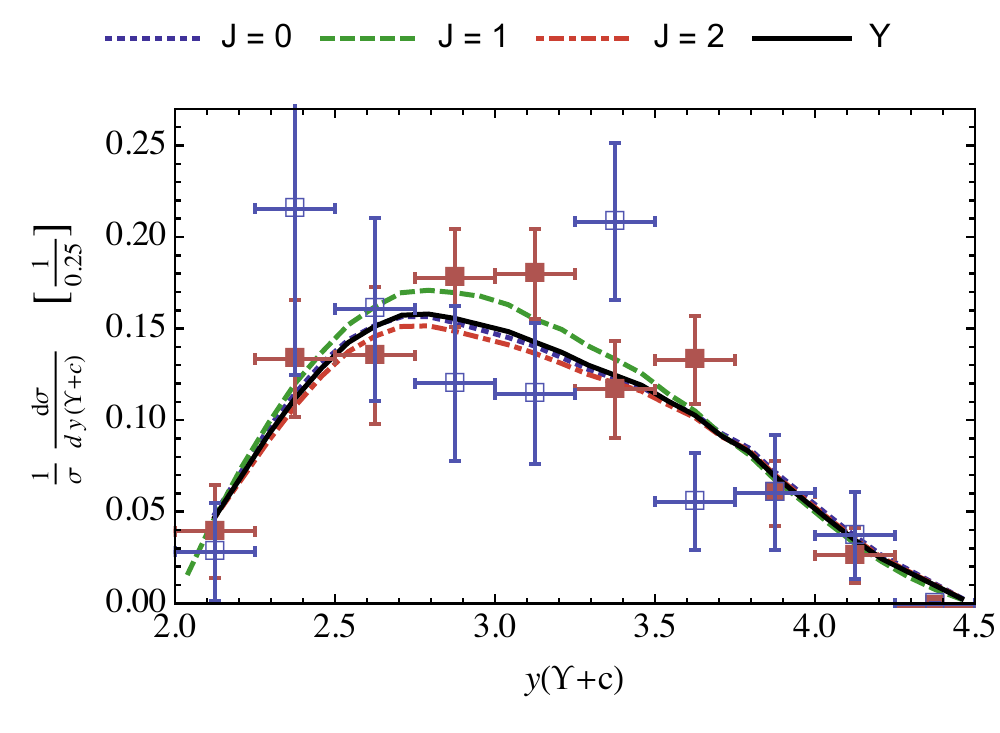}
\end{tabular*}
\caption{ \label{fig_distr_for_Y_3} Normalized cross section distributions for transverse momentum and rapidity of $\Upsilon + c$ system obtained within the SPS model. The definitions are the same as for Fig.~\ref{fig_distr_for_Y_1}.}
\end{figure*}

Fig.~\ref{fig_distr_for_Y_1} shows the normalized cross section distributions for transverse momentum and rapidity of $\Upsilon(1S)$ produced in radiative $\chi_b$ decays, $c$-quark and system of $\Upsilon(1S) + c$. As we have already noted, for the simplicity we assume that $c$-quark hadronizes into D meson with almost same $p_T$ and $y$. The experimental points from the LHCb \cite{Aaij:2015wpa} both for $\Upsilon(1S)\Dmeson^0$ and $\Upsilon(1S)\Dmeson^+$ final states are shown. 
Figs.~\ref{fig_distr_for_Y_2} and \ref{fig_distr_for_Y_3} shows same distributions for $c$-quark and $\Upsilon+c$ system. These figures show that the experimental data on distributions for $p_T$ and $y$  is poorly distinguishable from SPS predictions, i.e. the feed-down from the $\chi_b + c\bar c$ produced via SPS process gives the same contribution into the distribution shape as a DPS mechanism.

\begin{figure*}[t]
\begin{tabular*}{\textwidth}{c @{\extracolsep{\fill}} c}
\includegraphics[width=0.45\textwidth]{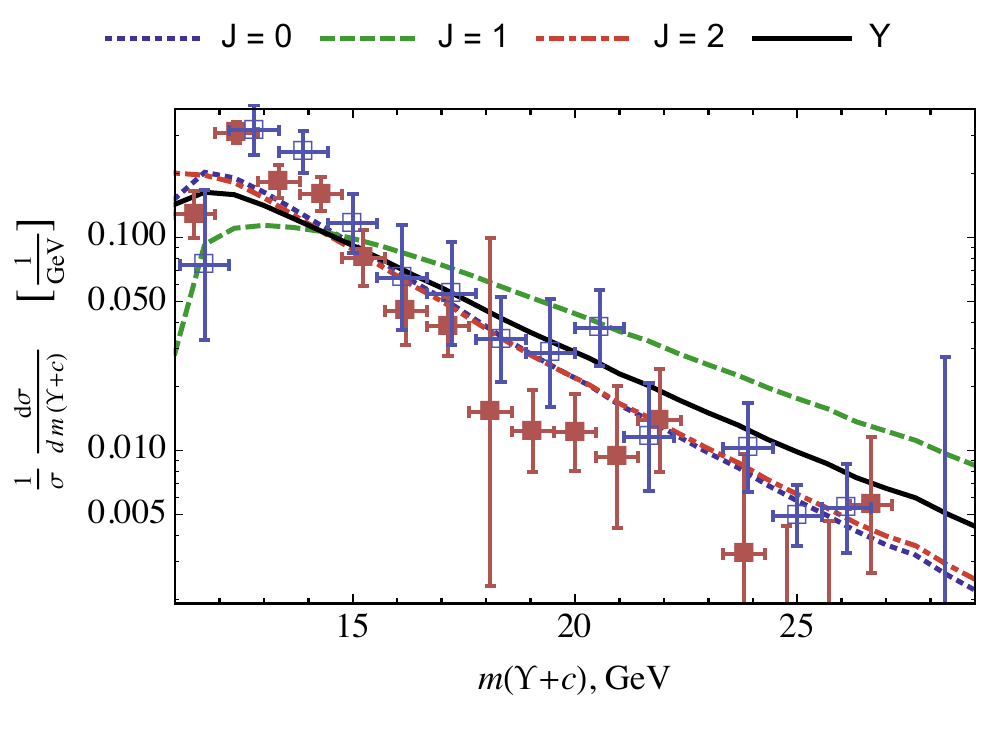} & \includegraphics[width=0.45\textwidth]{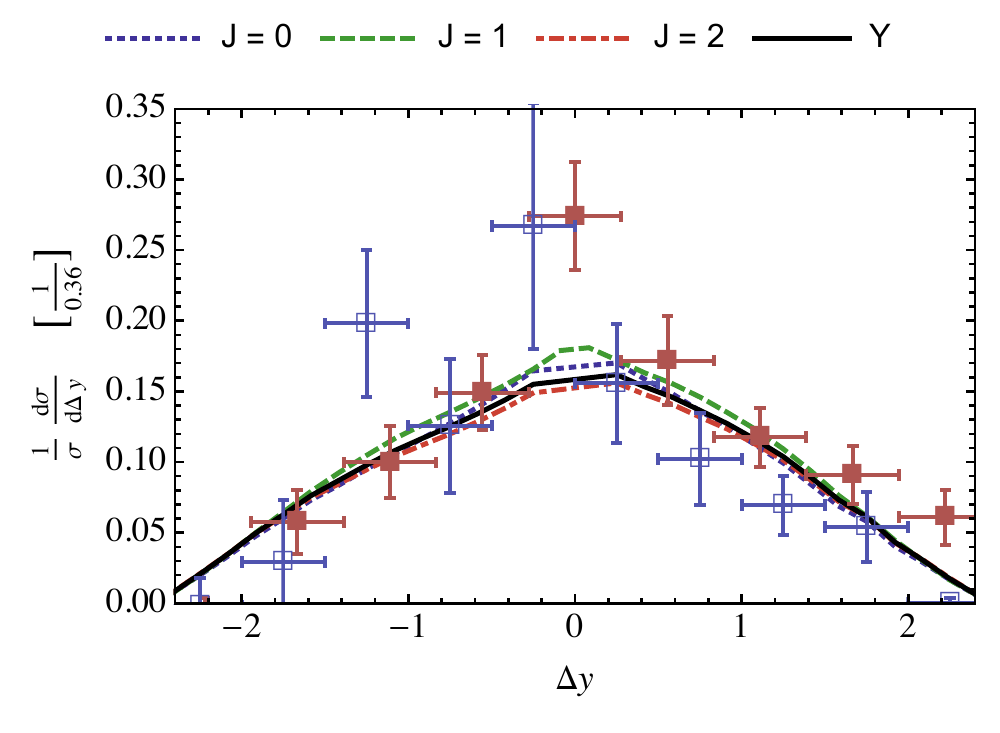}
\end{tabular*}
\caption{ \label{fig_distr_correlations_dy} Normalized cross section distributions for invariant mass of $\Upsilon$+$c$-quark system (left) and $|\triangle y|$ (right) obtained within the SPS model. The definitions are the same as for Fig.~\ref{fig_distr_for_Y_1}. }
\end{figure*}

Fig.~\ref{fig_distr_correlations_dy} shows distributions of the invariant mass of $\Upsilon + c$ system and for the rapidity asymmetry:
$$
\triangle y = y^{\Upsilon(1S)}  - y^{c\mbox{\scriptsize-quark}}.
$$
Again, experimental points are well fitted by the SPS curves.

\begin{figure*}[t]
\begin{tabular*}{\textwidth}{c @{\extracolsep{\fill}} c}
\includegraphics[width=0.45\textwidth]{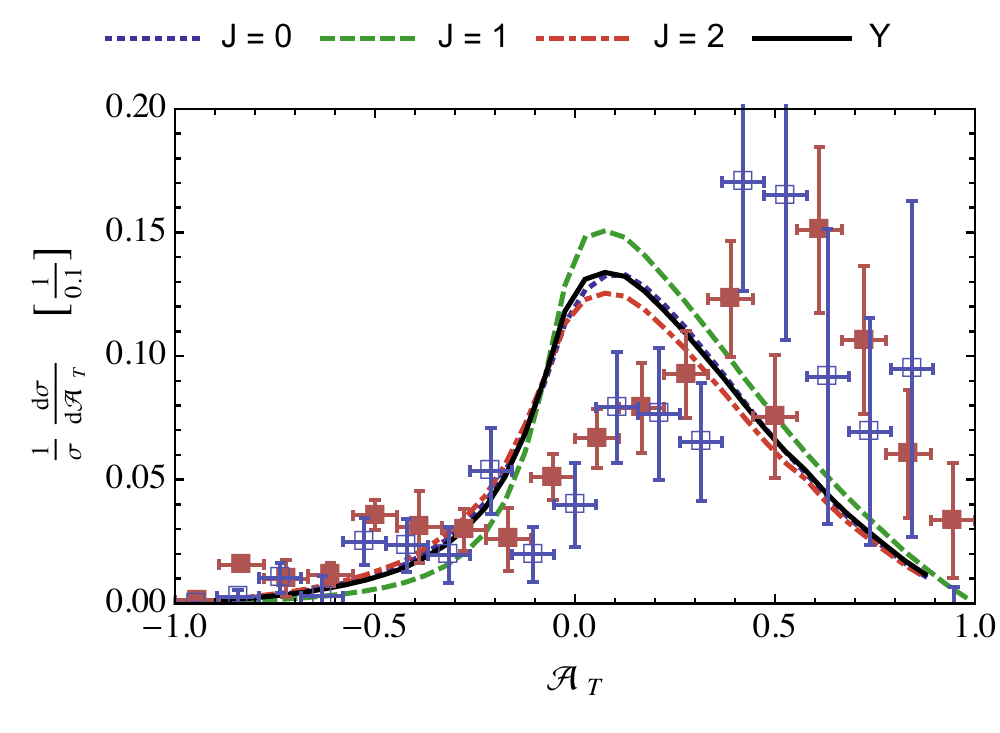} & \includegraphics[width=0.45\textwidth]{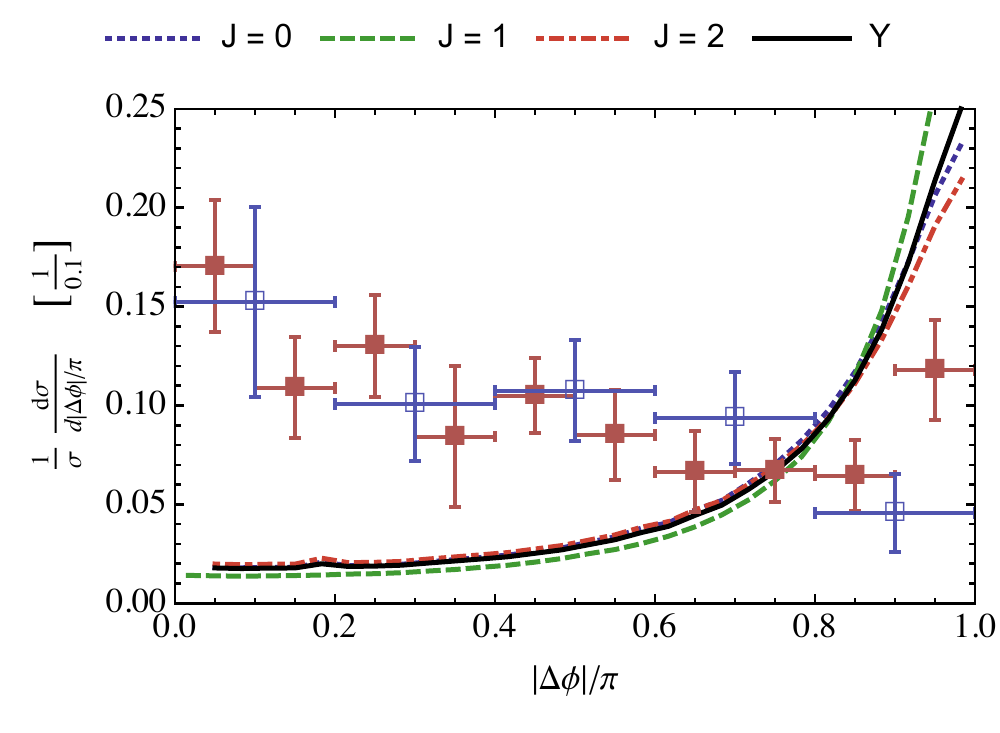}
\end{tabular*}
\caption{ \label{fig_distr_correlations_at} Normalized cross section distributions for $\mathcal A_T$ (left) and for $|\triangle \phi|/\pi$ (right) obtained within the SPS model. The definitions are the same as for Fig.~\ref{fig_distr_for_Y_1}. }
\end{figure*}

Finally, the results for the $p_T$ and azimuthal asymmetries:
\begin{eqnarray*}
\mathcal A_T \,=\, \frac{p_T^{\Upsilon(1S)}  - p_T^{c\mbox{\scriptsize-quark}}}{p_T^{\Upsilon(1S)}  + p_T^{c\mbox{\scriptsize-quark}}}, \quad
\left| \triangle \phi \right| \,=\, \left| \,  \phi^{\Upsilon(1S)}  - \phi^{c\mbox{\scriptsize-quark}} \, \right|
\end{eqnarray*}
are shown in Fig.~\ref{fig_distr_correlations_at}. In this figure we see the difference between SPS predictions and experimental data. In the case of $\mathcal A_T$, the difference is not large, especially when recall that the fragmentation $c\to \Dmeson^{0,+}$ (which we did not take into account) can slightly shift right our theoretical predictions. 

On the other hand, the azimuthal asymmetry shows significant difference between the experimental points and SPS predictions. It is clear that in the SPS model $\phi^{\Upsilon(1S)} - \phi^{c\bar c}  = \pi$ and if $c \bar c$ pair is highly correlated $\phi^{c}\approx \phi^{\bar c}$, we have $\left| \triangle \phi \right|$ strongly increasing at  $\pi$. Thus, $\left| \triangle \phi \right|$  is the most sensitive variable to the difference between SPS and DPS models. However, we admit that SPS predictions for $\left| \triangle \phi \right|$ may change significantly at the Next-to-Leading-Order, where the correlation between $c \bar c$ pair is not so crucial. Given that as discussed in the Introduction the NLO cross section may be comparable with the LO, this source may reduce the difference.

\section{Discussion}

In this paper we have studied the associated production of the $\chi_b$ mesons and open charm within SPS approach. We considered the feed-down of these processes to the $\Upsilon$ and open charm and found that it is about $(0.6 \div 1.5)\%$ of the experimental value for the total $\Upsilon + c\bar c$ found by the LHCb \cite{Aaij:2015wpa}. This may be a strong indication of the fact that other channels like NLO contributions, hidden charm and beauty hadronic components, DPS etc.~may come to the fore. What is more intriguing is that the shapes of the cross section distributions that we have found are well fit into the data obtained by the LHCb, except the azimuthal asymmetry which tends to be almost flat. We would like to stress that the strong peak in the SPS distribution may become flatten if consider real radiation in the NLO as it was seen for the paired production of $J/\psi$ \cite{Sun:2014gca}. Summarising, we admit that the precise account of the SPS mechanism should be done in order to correctly extract  contributions from the other possible channels.

\section*{Acknowledgments}
We would like to thank I. Belyaev and A. Berezhnoy for fruitful discussions. The work was supported by the RFBR grant \#14-02-00096A and partially by the RFBR grant \#15-02-03244A. The work of S. V. P. was supported by the RFBR grant \#16-32-60017.

\appendix

\section{Helicity projection method}
\label{app_calc}
The general form of the matrix element corresponding for the process \eqref{eq_hard_process} can be written as 
\begin{equation}
\mathcal A = \bar b(p_1) \, \mathcal M_{\mu\nu} \, b(p_2) \,\, \epsilon^\mu(k_1) \epsilon^\nu(k_2),
\end{equation}
where $\bar b(p_1)$ and $b(p_2)$ are spinors of final $b\bar b$ quark-antiquark pair and the dependence on gluon polarisation vectors written explicitly. Note, that we are working in QCD Lorentz gauge and diagrams in Fig.~\ref{fig_diagrams} contain 3-gluon vertex, so in general we need to add contributions from Faddeev-Popov ghosts as well (or work in another, e.g. axial, gauge). In order to avoid this complication we will use explicit values for gluon polarization vectors.

Projection onto a state with $S = 1$ can be done using a well known spin-projection operator:
\begin{equation*}
\Pi_{1,S_z} = \frac{1}{\sqrt{8 m_b^3}} \left( \frac{\slashed{P}}{2}+\slashed{q} + m_b \right) \slashed{\epsilon}(S_z) \left( \frac{\slashed{P}}{2}-\slashed{q} - m_b \right),
\end{equation*}
where $q$ is a relative momentum of $b \bar b$ pair and ${\epsilon}(S_z)$ is a spin polarization vector. In order to project $b \bar b$ pair onto state with fixed $L$, one need to consider the first non vanishing term in $q$ expansion, which in the case of $P$-wave is the second one, since $L=1$ wave function vanishes at the origin. Combining all together and restoring appropriate normalization factors we have for the amplitude:
\begin{multline}
\mathcal A(S_z, L_z) = -i\,R'(0)\,\sqrt{\frac{3}{4\pi}} \, \, \frac{1}{\sqrt{8 m_b^3}} \\
 \,\,\frac{d}{ d q_\alpha} \mbox{Tr}\left[\mathcal M_{\mu\nu} \, \left( \frac{\slashed{P}}{2}+\slashed{q} + m_b \right) \gamma_\beta \left( \frac{\slashed{P}}{2}-\slashed{q} - m_b \right) \right]_{q = 0} \\
 \epsilon^\alpha(L_z) \epsilon^\beta(S_z) \,\, \epsilon^\mu(k_1) \epsilon^\nu(k_2),
\end{multline}
Final projection onto $^3P_J$ states can be done using Clebsch-Gordan coefficients:
 \begin{equation}
 \label{eq_clebsch_gordan}
 \mathcal A(J_z) = \sum \langle S,S_z, L, L_z; J, J_z \rangle\,  \mathcal A(S_z, L_z)
 \end{equation}
 with
 \begin{eqnarray*}
&&   \sum \langle S,S_z, L, L_z; J, J_z \rangle\, \epsilon^\alpha(L_z) \epsilon^\beta(S_z) = P^{\alpha\beta},
\end{eqnarray*}
where each $P^{\alpha\beta}$ can be expressed in terms of quarkonia momentum and polarisation.
However, as we have found, summation over polarizations in the squared matrix element leads to an enormously huge expressions even for numerical processing. So we have proceed as follows. We chose a basis of three independent vectors of the following form:
\begin{eqnarray}
\label{eq_eps_x}
&&\epsilon^{(x)}_\alpha = c_1\, P_\alpha + c_2 (p_{1\alpha} + p_{2\alpha})\\
\label{eq_eps_y}
&&\epsilon^{(y)}_\alpha = c_3\, P_\alpha + c_4 (p_{1\alpha} + p_{2\alpha}) + c_5 k_{1\alpha}\\
\label{eq_eps_z}
&&\epsilon^{(z)}_\alpha = c_6\, \varepsilon_{\alpha\beta\mu\nu} \, k_1^\beta \, k_2^\mu \, (p_1^\nu + p_2^\nu)
\end{eqnarray}
where the coefficients are uniquely determined from the following equations:
\begin{eqnarray*}
&&\epsilon^{(y)}_\alpha \, \epsilon^{(x)}_\alpha = 0, \quad \epsilon^{(z)}_\alpha \, \epsilon^{(x)}_\alpha = 0, \quad \epsilon^{(x)}_\alpha \, \epsilon^{(y)}_\alpha = 0,\\
&&P_\alpha \, \epsilon^{(x,y,z)}_\alpha = 0,\quad \epsilon^{(x,y,z)}_\alpha \, \epsilon^{(x,y,z)}_\alpha = -1.
\end{eqnarray*}
With this notation $\epsilon^{(\pm)}_\alpha = \left. \left( \epsilon^{(x)}_\alpha \pm i \epsilon^{(y)}_\alpha \right) \right/ \sqrt{2}$ and $\epsilon^{(0)}_\alpha =\epsilon^{(z)}_\alpha$. The similar approach can be used for gluon polarizations.

Now, substituting $\epsilon(S_z)$ and $\epsilon(L_z)$ with one of (\ref{eq_eps_x}-\ref{eq_eps_z}) and doing the sane for gluon polarizations as well, we are able to calculate all 36 possible amplitudes. The major advantage of such an approach is that once all 36 amplitudes are calculated, we can obtain all helicity amplitudes for all values of total spin $J$ by simple summation using \eqref{eq_clebsch_gordan}.

Further simplification of the amplitudes can be performed if one observe that all spinor brackets coming from $c\bar c$ pair can be reduced to a set of just a few ones:
$$
\begin{array}{lll}
S_1 = \bar u v,  &S_2 = \bar u \gamma_5 v, & S_{3\mu} = \bar u \gamma_\mu v,\\
 S_{4\mu} = \bar u \gamma_5 \gamma_\mu v, & S_{5\mu\nu} = \bar u \gamma_\mu \gamma_\nu v, &S_{6\mu\nu} = \bar u \gamma_5 \gamma_\mu \gamma_\nu v.
\end{array}
$$
Contracting these structures with all possible momenta ($S_{3\mu} k_1^\mu$, $S_{3\mu} k_2^\mu$ etc.), we have 22 distinct Lorentz structures $L_i$ involving spinors. Thus each amplitude and matrix element can be written as
\begin{gather}
\label{eq_amp2_lor}
\mathcal A = \sum_{i=1}^{22} \mathcal A_i  L_i, \quad | \mathcal A |^2 = \sum_{i=1}^{22}\sum_{j=1}^{22}  \left(\mathcal A_i \mathcal A^*_j\right)  \left(L_i L^*_j\right),
\end{gather}
where all $\mathcal A_i$ are scalars free of spinors (and thus can be easily calculated numerically) and $L_i$ are all possible scalar combinations with spinors. When squaring the amplitude and summing over $c$-quark and $\bar c$-antiquark polarizations products like $\left(L_i L^*_j\right)$ will transform into a simple traces which can be precalculated. 

The final algorithm of the cross section calculation was the following. All $\mathcal A_i$ were calculated analytically using helicity projection method described above, all traces $\left(L_i L^*_j\right)$ were also calculated analytically. Having this done, equation \eqref{eq_amp2_lor} was used to obtain squared matrix element for each kinematical point. With this technique we have obtained a major performance boost (for example calculation of $10^6$ hadronic events takes just a few minutes on a standard laptop running in one processor thread).

Analytical calculations were performed using Redberry computer algebra system \cite{Bolotin:2013qgr}. Explicit analytical expressions are a bit tedious for pasting here, but they are available on request. The source code for the analytical part of calculation can be found at \href{http://github.com/PoslavskySV/pairedchi}{http://github.com/PoslavskySV/pairedchi} and for the numerical part at \href{http://bitbucket.org/ihep/chibcc}{http://bitbucket.org/ihep/chibcc}.

\bibliographystyle{apsrev4-1}
%

%\bibliography{references}

\end{document}